\documentclass[aps,prl,
preprint,
showpacs,
floatfix,
]{revtex4-1}
\usepackage{bm}
\usepackage{amsmath}    
\usepackage{graphicx}   
\usepackage[hang]{subfigure}

\begin{document}
\title {Magnetic Sensors based on Long Josephson Tunnel Junctions \\An Alternative to dc-SQUIDs}
\author{Roberto Monaco}
\affiliation{Istituto di Cibernetica del CNR, Comprensorio Olivetti, 80078 Pozzuoli, Italy \\and Facolt$\grave{\rm a}$ di Scienze, Universit$\grave{\rm a}$ di Salerno, 84084 Fisciano, Italy}\email
{r.monaco@cib.na.cnr.it}
\date{\today}
\begin{abstract}
The properties of Josephson devices are strongly affected by geometrical effects. A loop-shaped superconducting electrode tightly couples a long Josephson tunnel junction with the surrounding electromagnetic field. Due to the fluxoid conservation, any change of the magnetic flux linked to the loop results in a variation of the shielding current circulating around the loop, which, in turn, affects the critical current of the Josephson junction. This method allows the realization of a novel family of robust superconducting devices (not based on the quantum interference) which can function as a general-purpose magnetic sensors. The best performance is accomplished without compromising the noise performance by employing an in-line-type junction few times longer than its Josephson penetration length. The linear (rather than periodic) response to magnetic flux changes over a wide range is just one of its several advantages compared to the most sensitive magnetic detectors currently available, namely the Superconducting Quantum Interference Devices (SQUID). We will also comment on the drawbacks of the proposed system and speculate on its noise properties. 
\end{abstract}

\maketitle
\tableofcontents
\section{Introduction}
A Josephson quantum interferometer, which in its conventional form is realized by a superconducting loop interrupted by two Josephson junctions, lies at the core of the most sensitive magnetic detectors currently available, namely the Superconducting Quantum Interference Devices (SQUIDs)\cite{handbook}. Its working principle is that a variation, $\Delta \Phi_e$, of the external flux linked to the loop, smaller than one half of the magnetic flux quantum, $\Phi_0\equiv h/2e$, produces a measurable modulation, $\Delta I_c$, of the junctions maximum zero-voltage current (critical current), $I_c$. Best operational conditions for the {\it bare} interferometer (uncoupled loop and unshunted junctions) require\cite{tesche} that the shielding parameter, $\beta_{L} \equiv 2 L_{int} I_{c} /\Phi_0$, is approximately equal to unity and the best performance is achieved with an optimal value of the interferometer inductance $L_{int} =O( 100\,$pH) resulting in an  average current responsivity $\left\langle I_\Phi \right\rangle \equiv  \Delta I_c / \Delta \Phi_e \sim 10-20\mu A/\Phi_0$, but only in a flux range as small as $\Phi_0/2$. It took more than 30 years to turn the Josephson interferometers into the nowadays SQUIDs that can measure, at their best, magnetic fields as low as several attotestlas or magnetic fluxes as small as $1\,\mu \Phi_0$.

\begin{figure}[b]
\centering
\includegraphics[width=8.0cm,height=5.0cm]{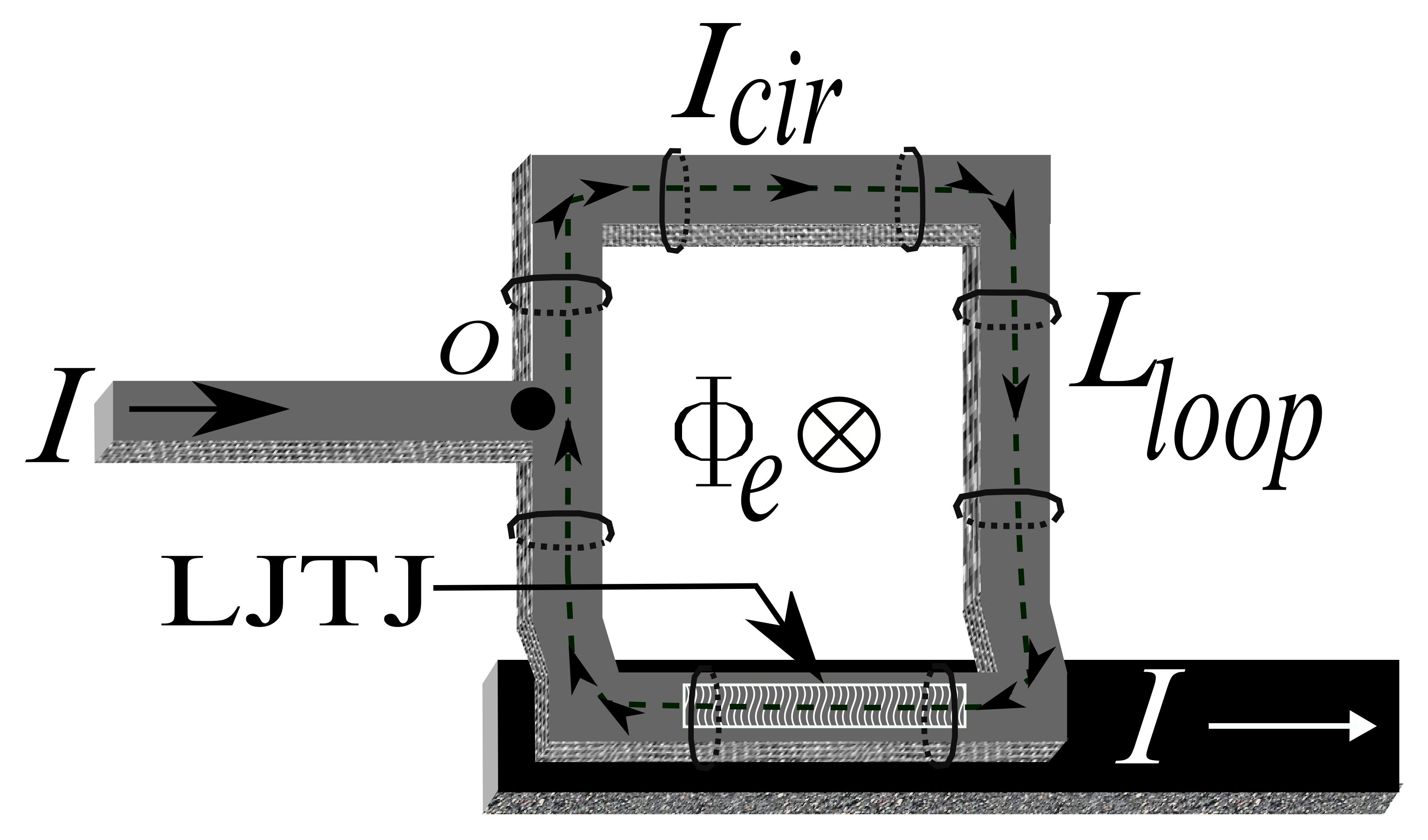}
\caption{3D sketch (not to scale) of the window-type in-line Josephson tunnel junction whose top electrode (in gray) is shaped as a rectangular loop. The base electrode is in black and the area of the tunneling insulating layer in between is hatched.}
\label{geometry}
\end{figure}

A constant flux-to-current transfer coefficient $I_\Phi \equiv  dI_c / d \Phi_e \approx 10\mu A/\Phi_0$ in a flux range of hundreds of flux quanta has been recently reported\cite{PRB12} using a Long Josephson Tunnel Junction (LJTJ) built on top of a superconducting loop, despite that the samples were not optimized for sensor applications. In its simplest configuration, a DOubly-Connected-Electrode Long Josephson Tunnel Junction (DOCELJTJ), consists of a LJTJ for which at least one electrode is shaped as a loop. This device is sketched in Fig.\ref{geometry} in which the junction bottom electrode is in black, while the top electrode - in the shape of a rectangular thin-film loop - is in gray and the tunneling area in between has a wavy hatched pattern. As it is generally accomplished with bare Josephson interferometers\cite{ketchen}, also for the DOCELJTJs the intrinsic flux sensitivity can be enhanced by several orders of magnitude, by efficiently coupling its loop to the secondary coil of a multi-turn input coil superconducting flux transformer. However, we will limit the interest to the bare DOCELJTJ and point out at some specific expedients that can be adopted to increase its flux signal-to-noise-ratio. A theoretical analysis of this system, corroborated by experiments, has been recently reported\cite{PRB12} in which the static sine-Gordon equation for a one-dimensional in-line LJTJ has been coupled to the quantization\cite{london} of the fluxoid in the doubly connected electrode; however, no mention was given in Ref.\cite{PRB12} about the the prospects of exploiting the DOCELJTJ as a magnetic sensor. In the next section we will review the working principles of this novel magnetic sensor and demonstrate that, with a proper design, it is competitive with the optimized Josephson interferometers. We will also comment on the drawbacks of the proposed system and speculate on its noise properties. Later on we will discuss its many advantages and point out at a few of its disadvantages, as well. Further discussion on noise limitation will be given in Section 3. In Section 4 the performance of a prototype device will be presented and commented upon. Finally, the conclusions will be drawn in the Section 5.

\section{The method}
Denoting with $L_{loop}$ the loop inductance and taking as positive the currents flowing clockwise, the DOCELJTJ working principle is the following. The internal magnetic flux, $\Phi_i$, within the loop is the sum of the externally applied  flux, $\Phi_e$, and the self-flux, $\Phi_s$, produced by the shielding current, $I_{cir}$, which circulates around the loop to restore the initial flux: $\Phi_i=\Phi_e+\Phi_s=n \Phi_0$. The last equality is a direct consequence of the London's fluxoid quantization law\cite{london}, where $n=0, \pm 1, \pm 2, ...$ is the number of flux quanta trapped in the loop at the time of its superconducting transition. Due to the fluxoid conservation\cite{mercereau}, any change in $\Phi_e$ corresponds to a variation of the circulating current, $I_{cir}\equiv \Phi_s/L_{loop}=(n\Phi_0- \Phi_e)/ L_{loop}$, which, in turn, alters the induced radial magnetic field, $H_{\rho} \propto I_{cir}$, at the loop surface, i.e., a superconducting loop acts as a flux-to-field transformer. If a magnetic field sensor is placed above (or below) part of the loop, it will thus detect the changes of the external magnetic flux, $\Phi_e$, linked to the loop. This remarkable property was first exploited by Pannetier {\it et al.}\cite{pannetier} who used high-sensitivity giant magnetoresistive sensors. However, within the context of superconducting thin films, the most natural (and sensitive) magnetic sensors are the Josephson junctions; in this case, any change in the external flux, $\Phi_e$, is measured by the variation, $\Delta I_c$, of the junction critical current. In essence, a DOCELJTJ is a flux-to-(critical) current transducer\cite{PRB09} which achieves its best performance with a one-dimensional Josephson tunnel junction whose width $W$ is smaller and whose length ${\rm{L}}$ is larger than the Josephson penetration length, $\lambda _J\equiv \sqrt{\Phi_0/ 2\pi \mu _{0}d_{e} J_{c}}$, setting the length unit of the LJTJ; $\mu _{0}$ is the vacuum magnetic permeability, $d_e $ the junction magnetic thickness\cite{wei} and $J_c$ the junction critical current density.  To bias the LJTJ, a dc current $I$ is fed into the loop at an arbitrary point $O$ and is inductively split in the two loop arms before crossing the LJTJ barrier; $I$ is taken out via the junction bottom electrode. With the bias current injected and extracted at the junction extremities we have the so called \textit{in-line} geometrical configuration treated in the pioneering works\cite{ferrelOS,stuehmbasa,radparvar85} on LJTJs soon after the discovery of the Josephson effect\cite{joseph}.  The junction critical current can be measured by standard time-of-fligth techniques\cite{fulton,castellano} based on AD conversion and peak detection with a resolution better than $1$ part in $10^3$, and it can be determined with at least one order of magnitude better accuracy, by measuring the switching current distributions\cite{russo}.

\noindent Upon the assumption that the junction width $W$ is larger than the magnetic thickness, $d_e$, of the Josephson sandwich, the radial magnetic field generated by the circulating current is $H_{\rho} = \Lambda_t I_{cir}/W$;  $\Lambda_t$ is the inductance per unit length of the junction top electrode normalized to the self-inductance per unit length, $\mathcal{L}_J=\mu_0 d_e/W_{}$, of the tunnel junction  seen as a two-conductor transmission line\cite{scott76,vanDuzer}. Of course, if the loop is formed by the bottom, rather than the top electrode, the normalized inductance per unit length of the junction bottom electrode, $\Lambda_b=1-\Lambda_t$, must replace $\Lambda_t$ in the above expression. Ideal symmetric junctions have $\Lambda_t=\Lambda_b=1/2$. For applications it is better to realize the loop with the top electrode which, typically, with respect to the bottom electrode, has a smaller width and so a larger inductance per unit length.

\begin{figure}[b]
\centering
\includegraphics[width=8.0cm,height=6cm]{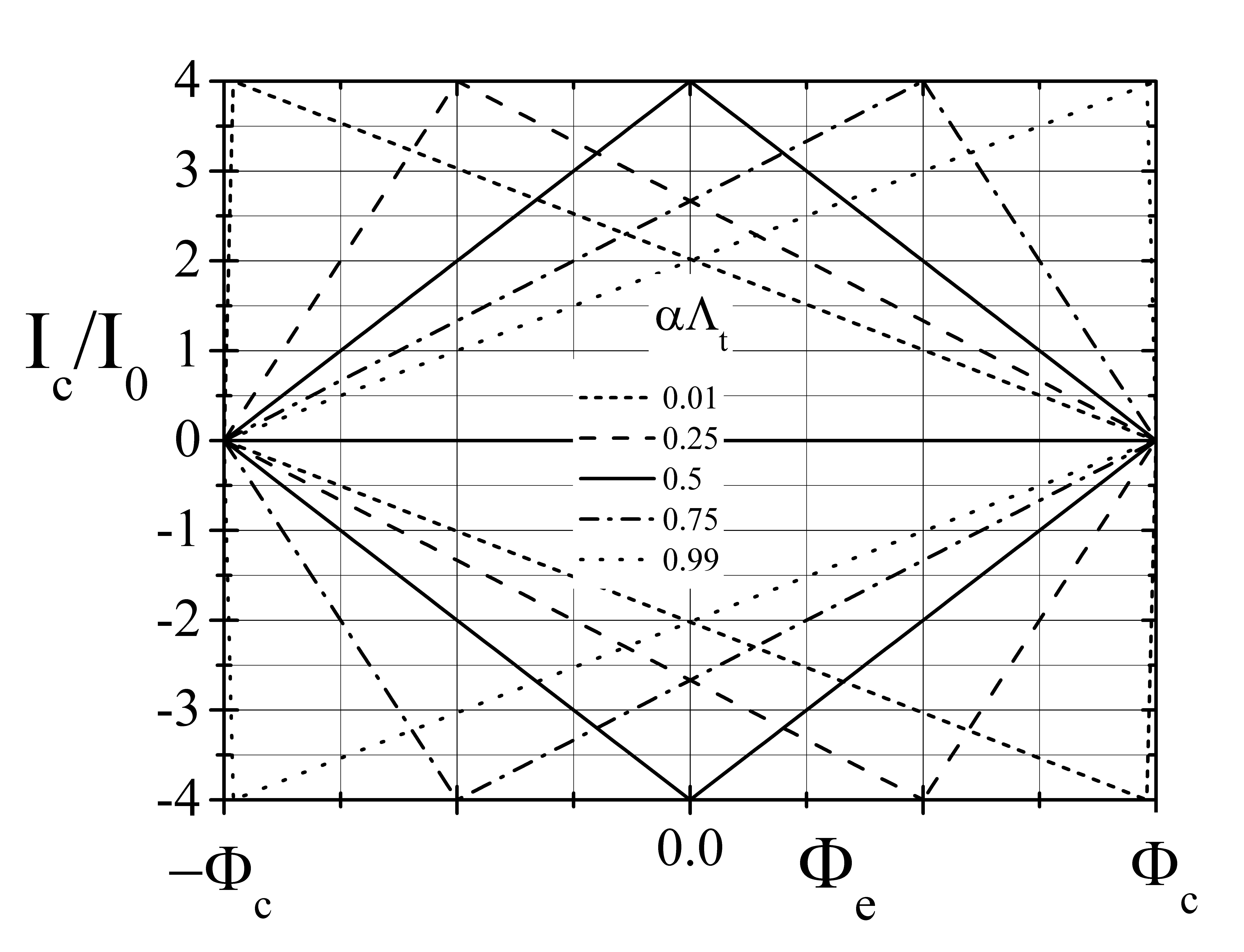}
\caption{Theoretical magnetic diffraction patterns $I_c(\Phi_e)$ for a very long noise-free in-line Josephson tunnel junction with different values of the balance parameter $0 \leq \alpha \Lambda_t \leq 1$. The critical current is normalized to $I_0 = J_c W_{} \lambda_J$ and the critical magnetic flux is $\Phi_c = 2 L_{loop} I_0/\Lambda_t$.} 
\label{mdpp}
\end{figure}

As reported by several authors\cite{stuehmbasa,vaglio}, the largest supercurrent carried by an in-line LJTJ is $4I_0$, where $I_0 \equiv J_c W \lambda_J$ is a characteristic junction current (generally from a fraction of a milliampere to a few milliamperes) and depends on the junction normalized length, $\ell\equiv {\rm{L}}/\lambda_J$, as $I_0(\ell)= I_0 \tanh \ell/2$. For $L>> \lambda_J$, strictly speaking in the limit $L \to \infty$, the threshold curves, $I_c(\Phi_e),$ in the Meissner regime and in the absence of noise, have been computed in Ref.\cite{PRB12} and are shown in Fig.~\ref{mdpp} for different values of the product $\alpha \Lambda_t$. Here $\alpha$ ($1-\alpha$) is a geometrical design parameter corresponding to the fraction of the bias current $I$ diverted into the right (left) arm of the loop; moving the current entry point $O$ along the loop, then the paths of least impedance change and $\alpha$ can take any value between $0$ and $1$. Since $\alpha$ and $\Lambda_t$ both belong to the interval $[0,1]$, so does their product. In Fig.2 the critical current $I_c$ is normalized to $I_0$ and the critical magnetic flux, $\Phi_c\equiv 2L_{loop}I_0/\Lambda_t$, is the theoretical flux value that fully suppresses the critical current. We have numerically computed the gradual crossover of $\Phi_c$ from short ($\ell\simeq 1$) to long ($\ell>2\pi$) junctions and it was found to be carefully described by the empirical relationship: $\Phi_c(\ell)= \Phi_c \coth \ell/\pi$.

\noindent At variance with a Josephson interferometer, a DOCELJTJ has a linear, rather than periodic, response to external flux changes; the wide range of linearity of the threshold curves is very attractive for the realization of high-dynamics sensors also in very noisy environments and makes the use of a flux-locked loop superfluous. The device sensitivity to flux changes is measured by the absolute slope $I_\Phi\equiv |dI_c/d\Phi_e|$:

\begin{equation}
\label{gain}
I_\Phi=\begin{cases} \frac{1}{\alpha L_{loop}}& \textrm{for} \quad -\Phi_c \leq \Phi_e \leq \Phi_{max}\\
\frac{\Lambda_t}{(1-\alpha \Lambda_t)L_{loop}} & \textrm{for} \quad \Phi_{max} \leq \Phi_e \leq \Phi_c
\end{cases}
\end{equation}


\noindent where $\Phi_{max}\equiv (2\alpha \Lambda_t -1)\Phi_c$. All-Niobium proof-of-principle DOCELJTJs were realized with $8\lambda_J$-long in-line junctions having the base electrode shaped as either a rectangular or annular closed path\cite{PRB12}. Their measurements accurately reproduced the theoretical predictions in Eqs.(\ref{gain}) in a wide flux range and for different $\alpha$ values. An intrinsic flux sensitivity larger than that of an optimized (bare) Josephson interferometer was attained with loop inductances less than $100\,$pH. It was also found that the device response can be tuned by an external magnetic field applied in the junction plane. In passing, we note that the product, $I_\Phi \Phi_c$, of the device sensitivity and dynamic range is independent on the loop inductance. Interestingly, the barrier electrical and geometrical parameters, such as the Josephson current density, $J_c$, and the junction width, $W_{}$, do not appear explicitly in Eqs.(\ref{gain}); the only requirement is that ${\rm{L}}>>\lambda_J$. For a finite length junction, $I_\Phi$ results to be proportional to the product $\tanh \ell/2 \times  \tanh \ell/\pi$ suggesting the use of very long junctions; however, for $\ell= 2\pi$, that product is already as large as $0.96$. In practical cases, it suffices to have $L \simeq 4-5 \lambda_J$ (and $W\leq \lambda_J/2$). By further decreasing the junction length, one would experience a progressive reduction of both the flux sensitivity and the linearity range. Incidentally, an enhanced stability against thermal fluctuations has been numerically reported\cite{pankratov} for the zero-voltage metastable states of (overlap-type) LJTJs with $L \simeq 5 \lambda_j$. We like to stress that, the DOCELJTJ loop not being interrupted by any Josephson element, the shielding parameter $\beta_L$ looses its importance which makes the junction critical current and the loop inductance independent design parameters. 

\begin{figure}[b]
\centering
\includegraphics[width=8cm]{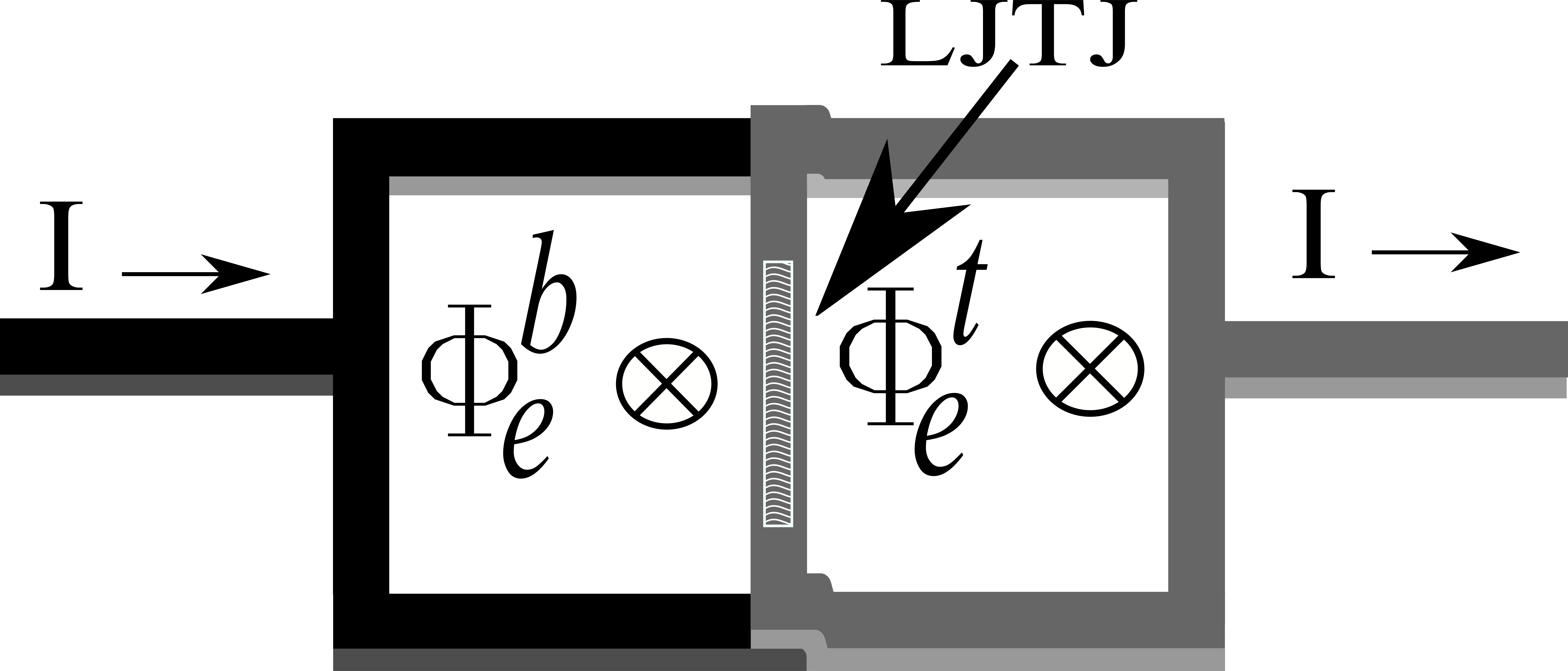}
\caption{3D sketch (not to scale) of a double loop DoCELJTJ. The base electrode is in black, the top electrode in gray and the tunneling insulating layer in between is white wavy hatched.}
\label{fig3}
\end{figure}

\noindent The threshold curves of Fig.~\ref{mdpp} are symmetric with respect to the inversion of both the supercurrent and the external flux, that is, $I_c^-(\Phi_e) = -I_c^+(-\Phi_e).$  Further, in the range $[-\Phi_{max},\Phi_{max}]$, the slope of the positive and negative critical currents, respectively, $I_{c}^+$ and $I_{c}^-$, is the same, i.e., any change in $\Phi_e$ modulates them in a concord fashion. This implies that the offset current, $I_c^{off} \equiv I_{c}^+ + I_{c}^-$, changes twice as fast and, $ I_{c}^+$ and $I_{c}^-$ being independent, its root mean square noise is $\sqrt{2}$ times larger than that of a single critical current; this enhances the signal-to-noise ratio by a factor $\sqrt{2}$. Notably, since the change with temperature of $I_{c}^+$ is numerically the same but opposite to that of $I_{c}^-$, one more advantage of the offset current, as compared to the single critical current, is its substantially reduced dependence on temperature.

\noindent Next, we consider the case when both electrodes are doubly connected as it is sketched in Fig.~\ref{fig3}. The theoretical analysis of a double loop device, not yet available, should account for two fluxoid quantizations (one in each loop) and the mutual magnetic interaction between the loops. Nevertheless, this double loop configuration is expected to be two times more sensitive to external flux changes: in fact, the shielding currents in the two loops circulate in opposite directions, but also on opposite sides with respect to the barrier, so that their respective radial magnetic fields add each other in the barrier plane.  In passing, we note that flux-focusing washer loops\cite{jaycox} or fractional-turn loops\cite{Zimmerman} made by many loops in parallel can also be usefully employed in the embodiment of a DOCELJTJ. Last but not least, if a parallel array of $N$ junctions is distributed along the loop perimeter, the resulting flux sensitivity will be $N$ times larger than that for just one junction and, at the same time, the signal-to-noise ratio is expected to increase by a factor $\sqrt{N}$.

\section{Noise properties}
The ultimate performance of any device depends on its noise. To estimate the value of the minimum detectable change of the external flux, it is important to know the spectral density, $S_\Phi$, of the flux noise generated under working conditions. Different sources contribute to the noise: we will only consider the intrinsic ones, since the extra noise induced by any eventual input circuitry has been already fully investigated in the context of the dc SQUIDs\cite{kleiner,koelle}. 

\noindent Any thermal fluctuation in the loop energy is felt as a magnetic flux noise $ \left\langle \Phi_n^2 \right\rangle=k_B T L_{loop}$, where $k_B$ is Boltzmann's constant and $T$ is the bath temperature; as for SQUIDs, ultra low noise applications demand small operating temperatures and loop inductances. However, the periodicity of any a Josephson interferometer would be completely wiped out by the noise, if $\left\langle \Phi_n^2 \right\rangle \geq \Phi_0^2/4$; the constraint $L_{int}< \Phi_0^2/4k_B T$ needed to observe quantum interference\cite{tesche} imposes $L_{int}< 15\,$nH, at liquid $He$ temperature and $L_{int}< 1\,$nH at liquid $N_2$ temperature. In the case of a DOCELJTJ, for a given operating temperature, is the requested measurement accuracy that determines the upper limit for the loop inductance.

\noindent The vast majority of dc SQUIDs are used at frequencies, $f$, below $1$kHz; if we focus on quasi-static or, at most, radio-frequency applications and assume thermal equilibrium with temperature $T >> hf/k_B$, we can disregard the shot noise due to the interaction of the current through the barrier and the electromagnetic field in the junction cavity\cite{likharev}. Furthermore, since the LJTJ is not shunted and operates in the zero-voltage state, we do not have to consider the flicker or $1/f$ noise; in addition, the Johnson noise generated in the large internal sub-gap resistance can be neglected in high-quality junctions operating well below their critical temperatures. At last, the only noise source intrinsic to the LJTJ is represented by the thermally induced escape from the zero-voltage state\cite{fulton} that manifests as a (critical) current noise with spectral density, $S_I$. The relative intensity of the thermal fluctuations is given by the dimensionless parameter $\Gamma\equiv k_B T/E_J$, where $E_J$ is the energy of the LJTJ. By adding the magnetostatic energy stored in and between the junction electrodes to the Josephson energy associated with the Cooper-pair tunneling current, it was found\cite{mc} that $E_J$ never exceeds $8E_0$, where $E_0\equiv \Phi_0 I_0/2\pi$ is the well-known fluxon rest energy; $E_0$ depends on the junction's electrical and geometrical parameters and represents its characteristic energy unit. Then, in our case, $\Gamma = \pi k_B T/4I_0 \Phi_0$, so that large $I_0$ values can be chosen to reduce the effects of the fluctuation; this fact is of paramount importance for the development of high-temperature sensors (with $I_0=0.5\,$mA, $\Gamma =O(10^{-3})$ at $T=77\,$K.) In Ref.\cite{castellano}, for bias currents near the critical current, the activation energy for (not very) long junctions was found to vary approximately as $(1-I/I_c)^{3/2}$, just as it does for short junctions\cite{fulton}, but its magnitude is scaled by a factor that depends on the junction normalized length and on the applied magnetic field and goes to zero at the critical field. Smaller activation energies result in a broader probability density for the escape from the zero-voltage state\cite{dino}, i.e., in a larger flux noise spectral density $S_\Phi=S_I/I_\Phi$.

The need for a cryogenic environment is the obvious drawback common to all superconducting devices. In our specific case, the main limitation is given by the smallest achievable size of the loop. Using the all-Niobium fabrication processes, high quality Josephson barriers can be attained\cite{highjc} with critical current densities as high as $10$kA/cm$^2$ corresponding to $\lambda_J \simeq 5 \mu$m; this is about the radius of the smallest useful ring for flux sensing applications. The high $J_c$ values required to reduce the Josephson depth, $\lambda_J$, at the same time, increase the characteristic current, $I_0$, so providing a broader linearity range, $\Phi_c$, and a smaller temperature parameter, $\Gamma$. The use a LJTJ is a drawback when an external shunt resistor would be used, as in the dc-SQUID, because here the involved resistance to be shunted to have a non hysteretic response is intrinsically lower than the one of the parallel connection of two (very) small junctions.

\section{The prototype}

\begin{figure}[htb]
\centering
\includegraphics[width=7cm]{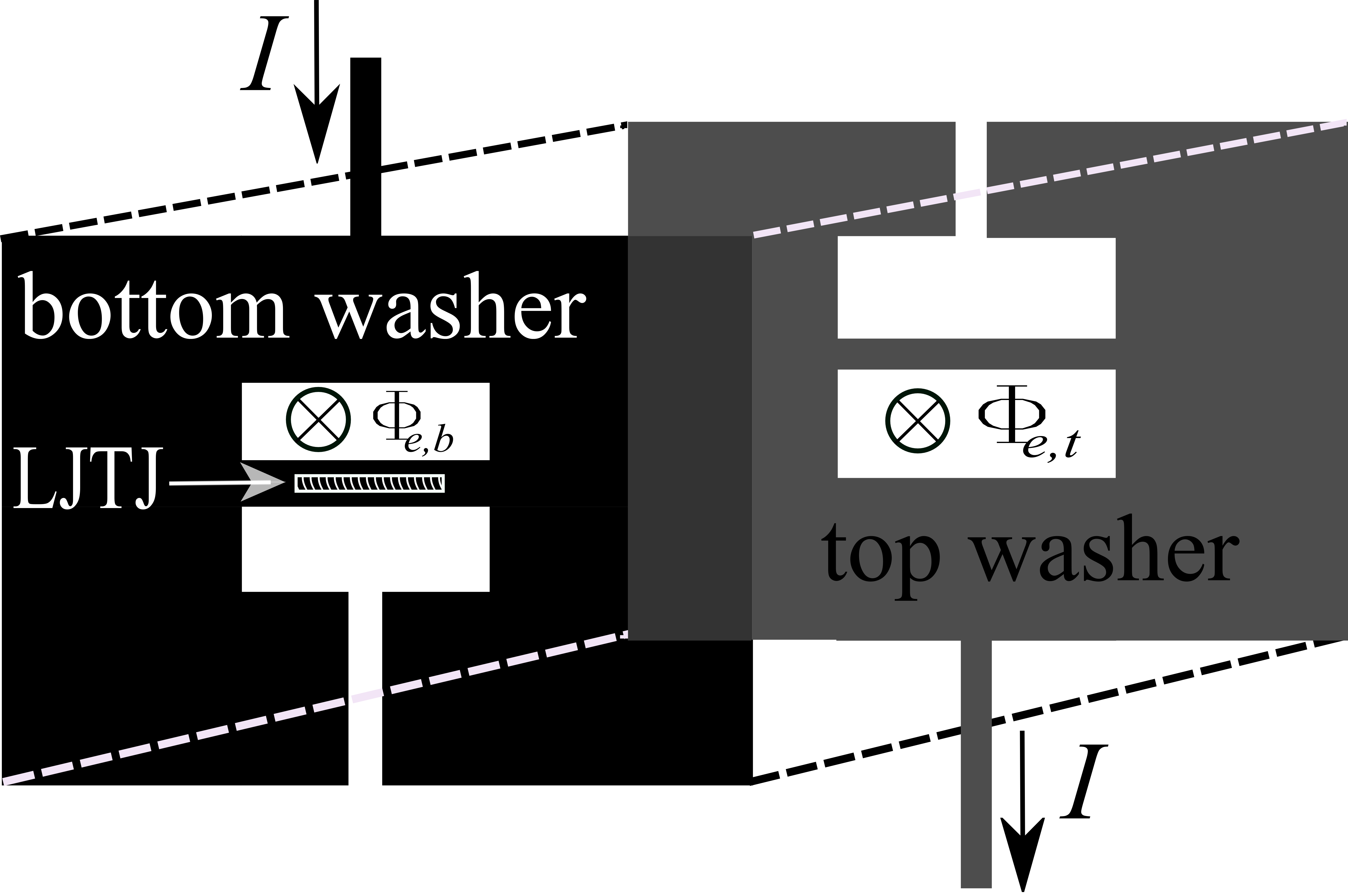}
\caption{Exploded diagram (not to scale) of a double loop device integrating two rectangular washer-type loops. The base electrode is in black, the top electrode in gray and the tunneling insulating layer in between is white wavy hatched.}
\label{fig4}
\end{figure}

Fig.~\ref{fig4} depicts the exploded view of a symmetric double loop DOCELJTJ realized with two rectangular flux-focusing washers\cite{jaycox} deformed in such a way that each acts as a ground plane for the other. The outer dimensions of each washer are $800\times 1000\, \mu m^2$, while the loop dimensions are $100\times 200\, \mu m^2$. Details on the samples fabrication and the experimental setup can be found in Ref.\cite{PRB12}. Here we will only point out the relevant electrical and geometrical parameters. We used high quality $Nb/Al-Al_{ox}/Nb$ LJTJs fabricated on silicon substrates using the trilayer technique in which the Josephson junction is realized in a window opened in a $200\,$nm thick $SiO_2$ insulator layer. The LJTJ had width $W_{}=1.5\,\mu$m and length ${\rm{L}}=100\,\mu$m and Josephson current density was $J_c\simeq 3.1\,$kA/cm$^2$ (at $T=4.2\,$K). In the junction area the bottom and top electrodes were, respectively, $10$ and $6\,\mu m$ wide and $100$ and $350\,nm$ thick. From the analysis of the experimental magnetic diffraction pattern in presence of a transverse magnetic field it is possible to derive peculiar system quantities such as the maximum critical current, $I_c^{max}=1.8\,$mA, the characteristic current, $I_0= 0.45\,$mA, and the symmetry parameter, $\alpha \Lambda_t= 0.55$. The magnetic current responsivity was found to be close to $2\mu A/nT$. The wires inside the cryoprobe were not filtered, so that the measurements were affected by a root mean square noise current $\left\langle I_n^2\right\rangle^{1/2}\sim 1\,\mu$A with an integration time of $0.1\,$s; therefore, the change in the static magnetic flux density that the bare DOCELJTJ could detect with a signal-to-noise ratio of one was $500pT$ in a $\pm 1\mu T$ range. We consider this as an encouraging figure of merit for a prototype sensor tested in an environment which is not meant for extremely low noise operations. The effective capture area of such a device is rather difficult to evaluate (as well as its self inductance); however, even assuming an underestimated value of $0.1\, mm^2$, we end up with a flux sensitivity better than $25m\Phi_0$ in a range of $\pm 100\Phi_0$. It is straightforward that a wider washer with a smaller loop area will drastically enhance the performance of the bare sensor. Furthermore, coupling the device to the secondary loop of a flux transformer would provide further orders of magnitudes improvements.

\section{Conclusions}
Long Josephson tunnel junctions were traditionally used to investigate the physics of non-linear phenomena\cite{barone}. In the last decade they have been employed to shed light on other fundamental concepts in physics such as the symmetry principles and how they are broken\cite{PRL06,PRB08,gordeeva}. In this paper we have discussed how one or more long Josephson tunnel junctions can be integrated with a superconducting loop to provide very large sensitivity to magnetic flux. The sensor principle is to capture the flux related to the field to be measured by the aim of a superconducting loop, perpendicular to the applied field. A supercurrent runs in the loop to avoid magnetic field penetration and to keep the superconductor in the Meissner state. If this loop is narrow, the circulating current density will become relatively high and will locally create a high magnetic field and a high density of field lines. A LJTJ placed above or below part of the loop will thus detect a local field through the changes of its critical current. The accuracy of the critical current measurements depends on the switching probability (or escape rate) caused by thermal noise.
We note that the LJTJ is sensitive to parallel fields, whereas the device (superconducting loop) is only sensitive to perpendicular fields. The physical property of superconductors which makes the operation of these devices possible is the quantization of the fluxoid associated with a closed loop of superconductor. We stress that the detection of magnetic flux with LJTJs is not based on the Josephson interference, but it only relies on the fluxoid conservation in the loop. This magnetometer combines ease of use, low noise, high dynamic performance and stability against thermal drifts. At the same time, it retains the advantages of high speed and low power inherent in Josephson devices. Its linear response is particularly advantageous also for the measurement of absolute flux and for noise thermometry. In addition, it is fully compatible with any present low- and high-T$_c$ thin film technology developed for the fabrication of Josephson junctions\cite{hilow}. Further, the demand on the external electronics is reduced, although a DOCELJTJ can benefit, in all respects, of the accessory circuitries (shunt resistors, input coil, modulation and feedback coil, flux transformer, tunable resonators and so on). In this light, the proposed device could conveniently replace the Josephson interferometer in the many dc SQUID sensors developed over the years to cover a wide range of applications (magnetometer, gradiometer, galvanometer, amplifier, etc.). Ultimately, the performances of any DOCELJTJ based device are intimately related to its intrinsic noise. Unfortunately, the noise properties of long Josephson junctions have not found an adequate interest in the Josephson community, and, in particular, the thermal effects in in-line LJTJs still remain an unexplored land. Further investigations have been planned to remedy this lack.



\begin{thebibliography}{50}

\bibitem{handbook} {\em The SQUID Handbook Fundamentals and Technology of SQUIDs and
SQUID Systems}\space, edited by J. Clarke and A. I. Braginski (Wiley-VCH
Verlag GmbH \& Co. KgaA, Weinheim, Germany, 2006), Vol 2; R. L.
Fagaly, {\it Rev. Sci. Instrum.} {\bf 77}, 101101 (2006).

\bibitem{tesche} C.D. Tesche and J. Clarke, {\it J. Low Temp. Phys.} {\bf 29}, 301 (1977). 

\bibitem{PRB12} R. Monaco, J. Mygind, and V.P. Koshelets, {\it Phys. Rev. B} {\bf 85}, 094514 (2012).

\bibitem{ketchen} M.B. Ketchen and J.M. Jaycox , {\it App. Phys.Letts.} {\bf 40}, 736 (1982).

\bibitem{london} F. London, {\it Phys. Rev.} {\bf 74}, 562 (1948).

\bibitem{mercereau} J. E. Mercereau and L. T. Crane, {\it Phys. Rev. Lett.}  {\bf 12}, 191 (1964). 

\bibitem{pannetier}  M. Pannetier, C. Fermon, G. Le Goff, J. Simola, E. Kerr, {\it Science}, {\bf 304}, 1648 (2004); M. Pannetier, C. Fermon, G. Le Goff, J. Simola, E. Kerr, M. Welling, and R.J. Wijngaarden, {\it IEEE Trans. on Appl. Supercond.} {\bf 15}, 892 (2005).

\bibitem{PRB09}  R. Monaco, J. Mygind, R.J. Rivers, and V.P. Koshelets, {\it Phys. Rev. B} {\bf 80}, 180501(R) (2009); John R. Kirtley and Francesco Tafuri, {\it Physics} {\bf 2}, 92 (2009).

\bibitem{wei} M. Weihnacht, {\it Phys. Status Solidi} {\bf 32}, K169 (1969).

\bibitem{ferrelOS}  R.A. Ferrel, and R.E. Prange, {\it Phys. Rev. Letts.}{\bf 10}, 479 (1963);
 C.S. Owen and D.J. Scalapino, {\it Phys. Rev.}{\bf 164}, 538 (1967).

\bibitem{stuehmbasa} D.L. Stuehm, and C.W. Wilmsem, {\it J. Appl. Phys.}{\bf 45}, 429 (1974); S. Basavaiah and R. F. Broom, {\it IEEE Trans. Magn.}{\bf MAG-11}, 759 (1975).

\bibitem{radparvar85} M. Radparvar and J. E. Nordman, {\it IEEE Trans. on Magn.} {\bf MAG-21}, 888 (1985).

\bibitem{joseph} B. D. Josephson, {\it Rev. Mod. Phys.} {\bf 36}, 216 (1964).

\bibitem{fulton} T. A. Fulton and L. N. Dunkleberger, {\it Phys. Rev. B} {\bf 9}, 4760 (1974).

\bibitem{castellano}  M.G. Castellano, G. Torrioli, C. Cosmelli, A. Costantini, F. Chiarello, P. Carelli, G. Rotoli, M. Cirillo, R.L. Kautz, {\it Phys. Rev. B} {\bf 54} 15417 (1996). 

\bibitem{russo} R. Russo, C. Granata, P. Walke, A. Vettoliere, E. Esposito, M. Russo, {\it J. Nanopart. Res.} {\bf 13}, 5661 (2011).

\bibitem{scott76} A.C. Scott, F.Y.F. Chu and S.A. Reible, {\it J. Appl. Phys.} {\bf 47}, 3272 (1976).

\bibitem{vanDuzer}  Theodore Van Duzer, Charles W. Turner, {\em Principles of Superconductive Devices and Circuits}\space (Prentice Hall- New Jersey, 2nd Edition, 1998).

\bibitem{vaglio} A. Barone, W.J. Johnson and R. Vaglio, {\it J. Appl. Phys.} {\bf 46}, 3628 (1975).

\bibitem{pankratov} K. G. Fedorov and A. L. Pankratov, {\it Phys. Rev. B} {\bf 76}, 024504 (2007).

\bibitem{jaycox} J.M. Jaycox and M.B. Ketchen, {\it IEEE Trans. Magn.} {\bf 17}, 403 (1981).

\bibitem{Zimmerman} J.E. Zimmerman, {\it Jou. Appl. Phys.} {\bf 42}, 4483 (1971).

\bibitem{kleiner} R. Kleiner, D. Koelle, F. Ludwig, and J. Clarke, {\it Proceedings of the IEEE} {\bf 92}, 1534 (2004).

\bibitem{koelle} D. Koelle, R. Kleiner, F. Ludwig, E. Dantsker and J. Clarke, {\it Rev. Mod. Phys.} {\bf 71}, 631 (1999).

\bibitem{likharev}  K.K. Likharev, {\em Dynamics of Josephson Junctions and Circuits}\space (Gordon \& Breach Science Publishers, London, 1984).

\bibitem{mc} A.C. Scott and W.J. Johnson, {\it Appl. Phys. Letts.}, {\bf 14}, 316 (1969); D.W. McLaughlin and A.C. Scott, {\it Phys. Rev. A}{\bf 18}, 1652 (1978).

\bibitem{dino} B. Ruggiero, C. Granata, E. Esposito, M. Russo and P. Silvestrini,
{\it Appl. Phys. Lett.} {\bf 75}, 121 (1999).

\bibitem{highjc} R. E. Miller, W. H. Mallison, A. W. Kleinsasser, K. A. Delin, and E. M. Macedo, {\it App. Phys.Letts.} {\bf 63}, 1423 (1993); H. Sugiyama, A. Fujimaki, and H. Hayakawa, {\it IEEE Trans. Appl. Superc.} {\bf 5}, 2739 (1995).

\bibitem{barone}  A. Barone and G. Patern\`o, {\em Physics and Applications of the Josephson Effect}\space (Wiley, New York, 1982).

\bibitem{PRL06} R. Monaco, J. Mygind,  M. Aaroe, R.J. Rivers and V.P. Koshelets, {\it Phys. Rev. Lett.} {\bf 96}, 180604 (2006). 

\bibitem{PRB08}  R. Monaco, M. Aaroe, J. Mygind, R.J. Rivers, and V.P. Koshelets, {\it Phys. Rev. B} {\bf 77}, 054509 (2008).

\bibitem{gordeeva}  A.V. Gordeeva and A.L. Pankratov, {\it Phys. Rev. B} {\bf 81}, 212504 (2010).

\bibitem{hilow} M. Gurvitch, M. A. Washington, and H. A. Huggins,  {\it App. Phys. Letts.} {\bf 42}, 472 (1983); H. Hilgenkamp and J. Mannhart, {\it Rev. Mod. Phys.} {\bf 74}, 485 (2002).

\end{thebibliography}
\end{document}